\documentclass[aps,prc,twocolumn,showpacs,floatfix,nofootinbib]{revtex4}
\usepackage{amsmath,graphicx}
\newcommand{\mean}[1]{\left\langle #1 \right\rangle}

\begin{document}

\preprint{}
\title{Relativistic viscous hydrodynamics for heavy-ion collisions:
A comparison between the Chapman-Enskog and Grad methods}

\author{Rajeev S. Bhalerao, Amaresh Jaiswal, Subrata Pal, and V. Sreekanth}
\affiliation {Tata Institute of Fundamental Research,
Homi Bhabha Road, Mumbai 400005, India}

\date{\today}

\begin{abstract}

Derivations of relativistic second-order dissipative hydrodynamic
equations have relied almost exclusively on the use of Grad's
14-moment approximation to write $f(x,p)$, the nonequilibrium
distribution function in the phase space. Here we consider an
alternative Chapman-Enskog-like method, which, unlike Grad's, involves
a small expansion parameter. We derive an expression for $f(x,p)$ to
second order in this parameter. We show analytically that while Grad's
method leads to the violation of the experimentally observed
$1/\sqrt{m_T}$ scaling of the longitudinal femtoscopic radii, the
alternative method does not exhibit such an unphysical behavior. We
compare numerical results for hadron transverse-momentum spectra and
femtoscopic radii obtained in these two methods, within the
one-dimensional scaling expansion scenario. Moreover, we demonstrate a
rapid convergence of the Chapman-Enskog-like expansion up to second
order. This leads to an expression for $\delta f(x,p)$ which provides
a better alternative to Grad's approximation for hydrodynamic modeling
of relativistic heavy-ion collisions.

\end{abstract}



\maketitle

\section{Introduction}

The standard model of relativistic heavy-ion collisions relies on
relativistic hydrodynamics to simulate the intermediate-stage
evolution of the high-energy-density fireball formed in these
collisions \cite{Heinz:2013th}. Recent simulations generally make use
of some version of the M\"uller-Israel-Stewart second-order theory of
causal dissipative hydrodynamics \cite{Israel:1979wp,Muronga:2003ta}.
Hydrodynamics has achieved remarkable success in explaining, for
example, the observed mass ordering of the elliptic flow
\cite{Adams:2005dq,Adcox:2004mh,Song:2013qma}, higher harmonics of the
azimuthal anisotropic flow \cite{Adare:2011tg,Chatrchyan:2013kba}, and
the ridge and shoulder structure in long-range rapidity correlations
\cite{ATLAS:2012at}. The recently measured correlators between event
planes of different harmonics \cite{Jia:2012sa} too can be understood
qualitatively within event-by-event hydrodynamics \cite{Qiu:2012uy}.
Notwithstanding these successes, the basic formulation of the
dissipative hydrodynamic equations continues to be an area of
considerable activity, largely because of the ambiguities arising due
to the variety of ways in which these equations can be derived
\cite{El:2009vj,Romatschke:2009im,Denicol:2012cn,Jaiswal:2012qm,Jaiswal:2013fc,Jaiswal:2013npa,Bazow:2013ifa}.
                                    
For a system that is out of equilibrium, the existence of
thermodynamic gradients results in thermodynamic forces, which give
rise to various transport phenomena. To quantify these nonequilibrium
effects, it is convenient to first specify the nonequilibrium
phase-space distribution function $f(x,p)$ and then calculate the
various transport coefficients. In the context of hydrodynamics, two
most commonly used methods to determine the form of the distribution
function close to local thermodynamic equilibrium are (1) Grad's
14-moment approximation \cite{Grad} and (2) the Chapman-Enskog method
\cite{Chapman}. Although both the methods involve expanding $f(x,p)$
around the equilibrium distribution function $f_0(x,p)$, there are
important differences.

In the relativistic version of Grad's 14-moment approximation, the
small deviation from equilibrium is usually approximated by means of a
Taylor-like series expansion in momenta truncated at quadratic order
\cite{Israel:1979wp,Romatschke:2009im}. Further, the 14 coefficients
in this expansion are assumed to be linear in dissipative
fluxes. However, it is not apparent why a power series in momenta
should be convergent and whether one is justified in making such
an ansatz, without a small expansion parameter.

The Chapman-Enskog method, on the other hand, aims at obtaining a
perturbative solution of the Boltzmann transport equation using the
Knudsen number (ratio of mean free path to a typical macroscopic
length) as a small expansion parameter. This is equivalent to making a
gradient expansion about the local equilibrium distribution function
\cite{deGroot}. This method of obtaining the form of the
nonequilibrium distribution function is consistent
\cite{Jaiswal:2013npa} with dissipative hydrodynamics, which is also
formulated as a gradient expansion.

The above two methods have been compared and shortcomings of Grad's
approximation have been pointed out in the literature
\cite{velasco,Calzetta:2013vma,Tsumura:2013uma}. In spite of these
shortcomings, the derivations of relativistic second-order dissipative
hydrodynamic equations, as well as particle-production prescriptions,
rely almost exclusively on Grad's approximation. The Chapman-Enskog
method, on the other hand, has seldom been employed in the
hydrodynamic modeling of the relativistic heavy-ion collisions. The
focus of the present work is to explore the applicability of the
latter method.

In this paper, the Boltzmann equation in the relaxation-time
approximation is solved iteratively, which results in a
Chapman-Enskog-like expansion of the nonequilibrium distribution
function. Truncating the expansion at the second order, we derive an
explicit expression for the viscous correction to the equilibrium
distribution function. We compare the hadronic spectra and
longitudinal Hanbury-Brown-Twiss (HBT) radii obtained using the form
of the viscous correction derived here and Grad's 14-moment
approximation, within a one-dimensional scaling expansion. We find that
at large transverse momenta, the present method yields smaller hadron
multiplicities. We also show analytically that while Grad's
approximation leads to the violation of the experimentally observed
$1/\sqrt{m_T}$ scaling of HBT radii
\cite{Beker:1994qv,Bearden:1998aq,Bearden:2001sy,Adcox:2002uc,LopezNoriega:2002tn}, 
the viscous correction obtained
here does not exhibit such unphysical behavior. Finally, we
demonstrate the rapid convergence of the Chapman-Enskog-like expansion
up to second order.


\section{Relativistic viscous hydrodynamics}

Within the framework of relativistic hydrodynamics, the variables that
characterize the macroscopic state of a system are the energy-momentum
tensor, $T^{\mu\nu}$, particle four-current, $N^\mu$, and entropy
four-current, $S^\mu$. The local conservation of net charge
($\partial_\mu N^\mu = 0$) and energy-momentum ($\partial_\mu
T^{\mu\nu}=0$) lead to the equations of motion of a relativistic
fluid, whereas the second law of thermodynamics requires
$\partial_\mu S^\mu \ge 0$. For a system with no net conserved
charges, hydrodynamic evolution is governed only by the conservation
equations for energy and momentum.

The energy-momentum tensor of a macroscopic system can be expressed in
terms of a single-particle phase-space distribution function and can
be tensor decomposed into hydrodynamic degrees of freedom \cite
{deGroot}. Here we restrict ourselves to a system of massless
particles (ultrarelativistic limit) for which the bulk viscosity
vanishes, leading to
\begin{align}\label{NTD}
T^{\mu\nu} &= \!\int\! dp \ p^\mu p^\nu\, f(x,p) = \epsilon u^\mu u^\nu 
- P\Delta ^{\mu \nu} + \pi^{\mu\nu}.
\end{align}
Here $dp\equiv g d{\bf p}/[(2 \pi)^3|\bf p|]$, where $g$ is the
degeneracy factor, $p^\mu$ is the particle four-momentum, and $f(x,p)$
is the phase-space distribution function. In the tensor
decomposition, $\epsilon$, $P$, and $\pi^{\mu\nu}$ are energy density,
thermodynamic pressure, and shear stress tensor, respectively.
The projection operator $\Delta^{\mu\nu}\equiv g^{\mu\nu}-u^\mu u^\nu$
is orthogonal to the hydrodynamic four-velocity $u^\mu$ defined in the
Landau frame: $T^{\mu\nu} u_\nu=\epsilon u^\mu$. The metric tensor is
Minkowskian, $g^{\mu\nu}\equiv\mathrm{diag}(+,-,-,-)$.

The evolution equations for $\epsilon$ and $u^\mu$,
\begin{align}\label{evol}
\dot\epsilon + (\epsilon+P)\theta - \pi^{\mu\nu}\nabla_{(\mu} u_{\nu)} &= 0,  \nonumber\\
(\epsilon+P)\dot u^\alpha - \nabla^\alpha P + \Delta^\alpha_\nu \partial_\mu \pi^{\mu\nu}  &= 0,
\end{align}
are obtained from the conservation of the energy-momentum tensor.  We
use the standard notation $\dot A\equiv u^\mu\partial_\mu A$ for
comoving derivative, $\theta\equiv \partial_\mu u^\mu$ for
expansion scalar, $A^{(\alpha}B^{\beta )}\equiv (A^\alpha B^\beta +
A^\beta B^\alpha)/2$ for symmetrization, and $\nabla^\alpha\equiv
\Delta^{\mu\alpha}\partial_\mu$ for spacelike derivatives. In the
ultrarelativistic limit, the equation of state relating energy density
and pressure is $\epsilon=3P\propto\beta^{-4}$. The inverse
temperature, $\beta\equiv1/T$, is determined by the Landau matching
condition $\epsilon=\epsilon_0$ where $\epsilon_0$ is the equilibrium
energy density. In this limit, the derivatives of $\beta$,
\begin{align}
\dot\beta &= \frac{\beta}{3}\theta - \frac{\beta}{12P}\pi^{\rho\gamma}\sigma_{\rho\gamma}, \label{evol1} \\
\nabla^\alpha\beta &= -\beta\dot u^\alpha - \frac{\beta}{4P} \Delta^\alpha_\rho \partial_\gamma \pi^{\rho\gamma}, \label{evol2}
\end{align}
can be obtained from Eq. (\ref{evol}), where
$\sigma^{\rho\gamma}\equiv\nabla^{(\rho}u^{\gamma)}-(\theta/3)\Delta^{\rho
  \gamma}$ is the velocity stress tensor \cite{Jaiswal:2013vta}. The
above identities are used later in the derivations of viscous
corrections to the distribution function and shear evolution equation.

For a system close to local thermodynamic equilibrium, the phase-space
distribution function can be written as $f=f_0+\delta f$, where the
deviation from equilibrium is assumed to be small $(\delta f\ll
f)$. Here $f_0$ represents the equilibrium distribution function of
massless Boltzmann particles at vanishing chemical potential,
$f_0=\exp(-\beta\,u\cdot p)$, where $u \cdot p \equiv u_\mu p^\mu$.
From Eq. (\ref {NTD}), the shear stress
tensor, $\pi^{\mu\nu}$, can be expressed in terms of the
nonequilibrium part of the distribution function, $\delta f$, as
\cite{Romatschke:2009im}
\begin{align}\label{FSE}
\pi^{\mu\nu} &= \Delta^{\mu\nu}_{\alpha\beta} \int dp \, p^\alpha p^\beta\, \delta f,
\end{align}
where $\Delta^{\mu\nu}_{\alpha\beta}\equiv
\Delta^{\mu}_{(\alpha}\Delta^{\nu}_{\beta)} -
(1/3)\Delta^{\mu\nu}\Delta_{\alpha\beta}$ is a traceless symmetric
projection operator orthogonal to $u^\mu$. To make further progress,
the form of $\delta f$ has to be determined. In the following, we
adopt a Chapman-Enskog-like expansion for the distribution function,
to obtain $\delta f$ order-by-order in gradients, by solving the
Boltzmann equation iteratively in the relaxation-time approximation.


\section{Chapman-Enskog expansion}

Determination of the nonequilibrium phase-space distribution function
is one of the central problems in statistical mechanics. This can be
achieved by solving a kinetic equation such as the Boltzmann
equation. The relativistic Boltzmann equation with the relaxation-time
approximation for the collision term is given by \cite
{Anderson_Witting},
\begin{equation}\label{RBE}
p^\mu\partial_\mu f = C[f] = 
-\left(u\!\cdot\! p\right) \frac{\delta f}{\tau_R},
\end{equation}
where $\tau_R$ is the relaxation time. We recall that the zeroth and
first moments of the collision term, $C[f]$, should vanish to ensure
the conservation of particle current and energy-momentum tensor \cite
{deGroot}. This requires that $\tau_R$ is independent of momenta, and
$u^\mu$ is defined in the Landau frame \cite
{Anderson_Witting}. Therefore, within the relaxation-time
approximation, Landau frame is mandatory and not a choice.
Momentum-dependent $\tau_R$ was considered in Ref.
\cite{Dusling:2009df} where the authors also studied the consequences of
different momentum dependencies of $\delta f$ for the heavy-ion
observables.

Exact solutions of the Boltzmann equation are possible only in rare
circumstances. The most common technique of generating an approximate
solution to the Boltzmann equation is the Chapman-Enskog expansion,
where the distribution function is expanded about its equilibrium
value in powers of space-time gradients \cite{Chapman}
\begin{equation}\label{CEE}
f = f_0 + \delta f, \quad \delta f= \delta f^{(1)} + \delta f^{(2)} + \cdots,
\end{equation}
where $\delta f^{(n)}$ is $n$th-order in derivatives. The Boltzmann
equation can be solved iteratively by rewriting Eq. (\ref{RBE}) in the
form $f=f_0-(\tau_R/u\cdot p)\,p^\mu\partial_\mu f$
\cite{Romatschke:2011qp,Jaiswal:2013npa,Teaney:2013gca}. We obtain
\begin{equation}\label{F1F2}
f_1 = f_0 -\frac{\tau_R}{u\cdot p} \, p^\mu \partial_\mu f_0, \quad 
f_2 = f_0 -\frac{\tau_R}{u\cdot p} \, p^\mu \partial_\mu f_1, ~~\, \cdots
\end{equation}
where $f_n=f_0+\delta f^{(1)}+\delta f^{(2)}+\cdots+\delta f^{(n)}$.
To first- and second-orders in derivatives, we have
\begin{align}
\delta f^{(1)} &= -\frac{\tau_R}{u\cdot p} \, p^\mu \partial_\mu f_0, \label{FOC} \\
\delta f^{(2)} &= \frac{\tau_R}{u\cdot p}p^\mu p^\nu\partial_\mu\Big(\frac{\tau_R}{u\cdot p} \partial_\nu f_0\Big). \label{SOC}
\end{align}
In the next section, the above expressions for $\delta f$ along with
Eq. (\ref{FSE}) are used in the derivation of the evolution
equation for the shear stress tensor.


\section{Viscous evolution equation}

In order to complete the set of hydrodynamic equations,
Eq. (\ref{evol}), we need to derive an expression for the shear stress
tensor, $\pi^{\mu\nu}$. The first-order expression for $\pi^{\mu\nu}$
can be obtained from Eq. (\ref{FSE}) using $\delta f = \delta f^{(1)}$
from Eq. (\ref {FOC}),
\begin{align}
\pi^{\mu\nu} &= \Delta^{\mu\nu}_{\alpha\beta}\int dp \ p^\alpha p^\beta \left(-\frac{\tau_R}{u \cdot p} \, p^\gamma \partial_\gamma\, f_0\right) . \label{FOSE}
\end{align}
Using Eqs. (\ref{evol1}) and (\ref{evol2}) and keeping only those
terms which are first-order in gradients, the integral in the above
equation reduces to
\begin{equation}\label{FOE}
\pi^{\mu\nu} = 2\tau_R\beta_\pi\sigma^{\mu\nu},
\end{equation}
where $\beta_\pi = 4P/5$ \cite{Jaiswal:2013npa}.

The second-order evolution equation for shear stress tensor can also
be obtained in a similar way by using $\delta f = \delta
f^{(1)}+\delta f^{(2)}$ from Eqs. (\ref {FOC}) and (\ref {SOC}) in
Eq. (\ref{FSE}). Performing the integrations and using
Eqs. (\ref{evol1}), (\ref{evol2}) and (\ref{FOE}), we get
\cite{Jaiswal:2013npa,Jaiswal:2013vta}
\begin{equation}\label{SOSHEAR}
\dot{\pi}^{\langle\mu\nu\rangle} \!+ \frac{\pi^{\mu\nu}}{\tau_R}\!= 
2\beta_{\pi}\sigma^{\mu\nu}
\!+2\pi_\gamma^{\langle\mu}\omega^{\nu\rangle\gamma}
\!-\frac{10}{7}\pi_\gamma^{\langle\mu}\sigma^{\nu\rangle\gamma} 
\!-\frac{4}{3}\pi^{\mu\nu}\theta,
\end{equation}
where $\omega^{\mu\nu}\equiv(\nabla^\mu u^\nu-\nabla^\nu u^\mu)/2$ is
the vorticity tensor, and we have used Eq. (\ref{FOE}). It is clear
from the form of the above equation that the relaxation time $\tau_R$
can be identified with the shear relaxation time $\tau_\pi$. By
comparing the first-order evolution Eq. (\ref {FOE}) with the
relativistic Navier-Stokes equation
$\pi^{\mu\nu}=2\eta\sigma^{\mu\nu}$, we obtain
$\tau_\pi=\eta/\beta_\pi$, where $\eta$ is the coefficient of shear
viscosity.

\section{Corrections to the distribution function}

In this section, we derive the expression for the nonequilibrium part
of the distribution function, $\delta f$, up to second order in
gradients of $u^\mu$. For this purpose, we employ Eqs. (\ref{FOC}) and
(\ref{SOC}), which were obtained using a Chapman-Enskog-like
expansion. We then recall the derivation of the standard Grad's
14-moment approximation for $\delta f$, and compare these two
expressions.

Using Eqs. (\ref {evol1}) and (\ref {evol2}) for the derivatives of
$\beta$, and Eq. (\ref {SOSHEAR}) for $\sigma^{\mu\nu}$, in Eqs.
(\ref{FOC}) and (\ref {SOC}), we arrive at the form of the
second-order viscous correction to the distribution function:
\begin{align}
\delta f \!=\ &  \frac{f_0\beta}{2\beta_\pi(u\!\cdot\!p)}\, p^\alpha p^\beta \pi_{\alpha\beta}
-\frac{f_0\beta}{\beta_\pi} \bigg[\frac{\tau_\pi}{u\!\cdot\!p}\, p^\alpha p^\beta \pi^\gamma_\alpha\, \omega_{\beta\gamma} \nonumber\\
&-\frac{5}{14\beta_\pi (u\!\cdot\!p)}\, p^\alpha p^\beta \pi^\gamma_\alpha\, \pi_{\beta\gamma}
+\frac{\tau_\pi}{3(u\!\cdot\!p)}\, p^\alpha p^\beta \pi_{\alpha\beta}\theta  \nonumber\\
&-\frac{6\tau_\pi}{5}\, p^\alpha\dot u^\beta\pi_{\alpha\beta}
+\frac{(u\!\cdot\!p)}{70\beta_\pi}\, \pi^{\alpha\beta}\pi_{\alpha\beta}
+\frac{\tau_\pi}{5}\, p^\alpha \left(\nabla^\beta\pi_{\alpha\beta}\right) \nonumber\\
&-\frac{3\tau_\pi}{(u\!\cdot\!p)^2}\, p^\alpha p^\beta p^\gamma \pi_{\alpha\beta}\dot u_\gamma
+\frac{\tau_\pi}{2(u\!\cdot\!p)^2}\, p^\alpha p^\beta p^\gamma \left(\nabla_\gamma\pi_{\alpha\beta}\right) \nonumber\\
&-\frac{\beta+(u\!\cdot\!p)^{-1}}{4(u\!\cdot\!p)^2\beta_\pi}\, \left(p^\alpha p^\beta \pi_{\alpha\beta}\right)^2\bigg]
+{\cal O}(\delta^3), \label{SOVC} \\ 
\equiv~& \delta f_1 + \delta f_2 +{\cal O}(\delta^3). \label{SOVCE}
\end{align}
The first term on the right-hand side of Eq. (\ref{SOVC}) corresponds
to the first-order correction, $\delta f_1$, whereas the terms within
square brackets are of second order, $\delta f_2$ (see Appendix A).
Note that $\delta f_1 \ne \delta f^{(1)}$ and $\delta f_2 \ne \delta
f^{(2)}$, due to the nonlinear nature of Eqs. (\ref {evol1}), (\ref
{evol2}), and (\ref{SOSHEAR}). It is straightforward to show that the
form of $\delta f$ in Eq. (\ref{SOVC}) is consistent with the definition 
of the shear stress tensor, Eq. (\ref{FSE}), and satisfies the matching
condition $\epsilon =\epsilon_0$ and the Landau frame definition
$u_\nu T^{\mu \nu} = \epsilon u^\mu$ \cite{deGroot}, i.e.,
\begin{equation}\label{checks}
\int dp\, (u\cdot p)^2\, \delta f = 0, \quad \int dp\, \Delta_{\mu\alpha}u_\beta\, p^\alpha p^\beta\, \delta f = 0,
\end{equation}
order-by-order in gradients (see Appendix A).

On the other hand, Grad's 14-moment approximation for $\delta f$ can
be obtained from a Taylor-like expansion in the powers of momenta
\cite{Israel:1979wp,Romatschke:2009im}
\begin{equation}
 \delta f_G = f_0 \left[\varepsilon(x)+\varepsilon_\alpha(x)p^\alpha+\varepsilon_{\alpha\beta}(x)p^\alpha p^\beta \right],
\end{equation}
where $\varepsilon$'s are the momentum-independent coefficients in the
expansion, which, however, may depend on thermodynamic and dissipative
quantities. For a system of massless particles with no net conserved
charges, i.e., in the absence of bulk viscosity and charge diffusion
current, the above equation reduces to
\begin{equation}
 \delta f_G = \frac{f_0 \beta^2}{10 \beta_\pi}\, p^\alpha p^\beta \pi_{\alpha\beta}, \label{CG}
\end{equation}
where the coefficient is obtained using Eq. (\ref{FSE}). We observe
that unlike Eq. (\ref{SOVC}) for the Chapman-Enskog case,
Eq. (\ref{CG}) for Grad's is linear in shear stress tensor. However,
it is important to note that both the forms of $\delta f$, i.e.,
$\delta f_1$ and $\delta f_G$, lead to identical evolution equations
for the shear stress tensor, Eq. (\ref{SOSHEAR}), with the same
coefficients \cite{Denicol:2012cn,Jaiswal:2013vta}.


\section{Bjorken scenario}
In order to model the hydrodynamical evolution of the matter formed in
the heavy-ion collision experiments, we use the Bjorken prescription
\cite{Bjorken:1982qr} for one-dimensional expansion. We consider
the evolution of a system of massless particles ($\epsilon=3P$) at
vanishing net baryon number density. In terms of the Milne coordinates
($\tau,r,\varphi,\eta_s$), where $\tau = \sqrt{t^2-z^2}$,
$r=\sqrt{x^2+y^2}$, $\varphi=\tan^{-1}(y/x)$, and
$\eta_s=\tanh^{-1}(z/t)$, and with $u^\mu=(1,0,0,0)$, evolution
equations for $\epsilon$ and $\Phi \equiv -\tau^2 \pi^{\eta_s \eta_s}$
become
\begin{align}
\frac{d\epsilon}{d\tau} &= -\frac{1}{\tau}\left(\epsilon + P -\Phi\right), \label{BED} \\
\frac{d\Phi}{d\tau} &= - \frac{\Phi}{\tau_\pi} + \beta_\pi\frac{4}{3\tau} - \lambda\frac{\Phi}{\tau}. \label{Bshear}
\end{align}
The transport coefficients appearing in the above equation reduce to \cite{Jaiswal:2013npa}
\begin{equation}\label{BTC}
\tau_\pi = \frac{\eta}{\beta_\pi}, \quad \beta_\pi = \frac{4P}{5}, \quad \lambda = \frac{38}{21}.
\end{equation}

In $(\tau,r,\varphi,\eta_s)$ coordinates, the components of particle
four-momenta are given by
\begin{align}
p^\tau &= m_T \cosh(y-\eta_s), \quad p^r = p_T \cos(\varphi_p- \varphi), \\
p^\varphi &= p_T\sin(\varphi_p-\varphi)/r, \quad p^{\eta_s} = m_T \sinh(y-\eta_s)/\tau, \nonumber
\end{align}
where $m_T^2=p_T^2+m^2$, $p_T$ is the transverse momentum, $y$ is the
particle rapidity, and $\varphi_p$ is the azimuthal angle in the momentum
space. We note that for the Bjorken expansion, $\theta=1/\tau$, $\dot
u^{\mu}=0$, $\omega^{\mu\nu}=0$ and $p_\mu d\Sigma^\mu = m_T
\cosh(y-\eta_s) \tau d\eta_s r dr d\varphi$. In this scenario, the
nonvanishing factors appearing in Eq. (\ref{SOVC}) reduce to $u\cdot
p = m_T \cosh(y-\eta_s)$, $\pi_{\alpha\beta}\pi^{\alpha\beta}=
3\Phi^2/2$, and
\begin{align}
p^{\alpha}p^{\beta} \pi_{\alpha\beta} &= \frac{\Phi}{2}\, p_T^2-\Phi\, m_T^2\, \sinh^2(y-\eta_s), \nonumber\\
p^{\alpha}p^{\beta} \pi^\gamma_\alpha \pi_{\gamma\beta} &= -\frac{\Phi^2}{4}\, p_T^2-\Phi^2\, m_T^2\, \sinh^2(y-\eta_s), \nonumber\\
p^\alpha p^\beta p^\gamma \nabla_\alpha\pi_{\beta\gamma} &= 2\, \frac{\Phi}{\tau}\, m_T^3\, \sinh^2(y-\eta_s)\cosh(y-\eta_s), \nonumber\\
p^\alpha\nabla^\beta\pi_{\alpha\beta} &= -\frac{\Phi}{\tau}\, m_T\, \cosh(y-\eta_s). \label{visc-fcts}
\end{align}

Within the framework of the relativistic hydrodynamics, observables
pertaining to heavy-ion collisions are influenced by viscosity in two
ways: first through the viscous hydrodynamic evolution of the system
and second through corrections to the particle production rate via the
nonequilibrium distribution function \cite{Teaney:2003kp}.
Hydrodynamic evolution and the nonequilibrium corrections to the
distribution function were considered in the previous sections; in the
following sections, we focus on two observables, namely
transverse-momentum spectra and HBT radii of hadrons.


\section{Hadronic spectra}

The hadron spectra can be obtained using the Cooper-Frye freezeout
prescription \cite{Cooper:1974mv}
\begin{equation}\label{CF}
\frac{dN}{d^2p_Tdy} = \frac{g}{(2\pi)^3} \int p_\mu d\Sigma^\mu f(x,p),
\end{equation}
where $p^\mu$ is the particle four-momentum, $d\Sigma^\mu$ represents
the element of the three-dimensional freezeout hypersurface, and
$f(x,p)$ represents the phase-space distribution function at
freezeout.

For the ideal freezeout case ($f=f_0$), we get
\begin{equation}\label{ICF}
\frac{dN^{(0)}}{d^2p_Tdy} = \frac{g}{4 \pi^3}\, m_T\, \tau\, A_\perp\, K_1,
\end{equation}
where $A_\perp$ denotes the transverse area of the overlap zone of
colliding nuclei and $K_n \equiv K_n(z_m)$ are the modified Bessel
functions of the second kind with argument $z_m\equiv
m_T/T$. In Eq. (\ref{ICF}) and hereafter, the hydrodynamical quantities such as
$T,~\tau,~\Phi,~P$, etc., correspond to their values at freezeout. The
expression for hadron production up to first order ($f=f_0+\delta
f_1$) is obtained as
\begin{equation}\label{FCF}
\frac{d N^{(1)}}{d^2p_Tdy} = \left[1+ \frac{\Phi}{4\beta_\pi z_m} \left\{z_p^2\, \frac{K_0}{K_1} - 2z_m \right\}\right]\frac{dN^{(0)}}{d^2p_Tdy},
\end{equation}
where $z_p\equiv p_T/T$. Here we have used the recurrence relation
$K_{n+1}(z)=2nK_n(z)/z+K_{n-1}(z)$. The derivation of the hadron
spectra up to second order, $dN^{(2)} / d^2p_Tdy$ (by setting
$f=f_0+\delta f_1+\delta f_2$), is presented in the Appendix B.

For comparison, we also present the result for hadron production
obtained using Grad's 14-moment approximation ($f=f_0 + \delta f_G$)
\cite{Teaney:2003kp,Bhalerao:2013aha}
\begin{equation}\label{GCF}
\frac{d N^{(G)}}{d^2p_Tdy} = \left[1+\frac{\Phi}{20\beta_\pi} \left\{z_p^2 - 2z_m \frac{K_2}{K_1}  \right\}\right]\frac{dN^{(0)}}{d^2p_Tdy}.
\end{equation}

 \begin{figure}[t]
 \begin{center}
 \includegraphics[scale=0.37]{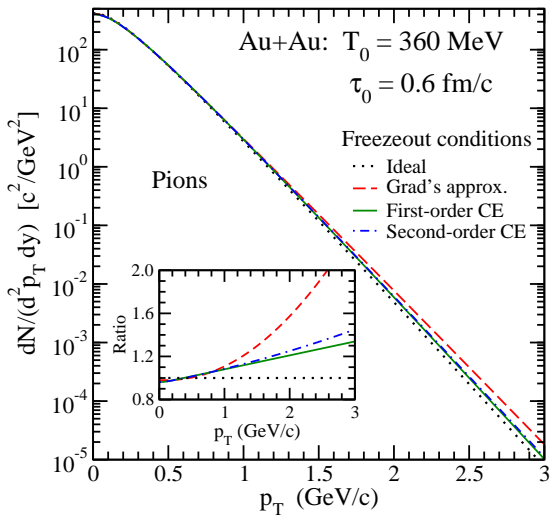}
 \end{center}
 \vspace{-0.4cm}
 \caption{(Color online) Pion spectra as a function of the transverse
   momentum $p_T$, obtained with the second-order hydrodynamic
   evolution, followed by freezeout in various scenarios: ideal, Grad's
   14-moment approximation, and first- and second-order
   Chapman-Enskog. Inset: Pion yields in the above four cases scaled
   by the corresponding values in the ideal case.}
 \label{spectra}
 \end{figure}

We solve the evolution equations (\ref{BED}) and (\ref{Bshear}) with
initial temperature $T_0=360$ MeV, time $\tau_0=0.6$ fm/$c$, and
isotropic pressure configuration $\Phi_0=0$, corresponding to central
($b=0$) Au-Au collisions at the Relativistic Heavy-Ion Collider. The
system is evolved with shear viscosity to entropy density ratio
$\eta/s=1/4\pi$ corresponding to the Kovtun-Son-Starinets (KSS) lower bound
\cite{Kovtun:2004de}, until the freezeout temperature $T=150$ MeV is
reached. In order to study the effects of the various forms of $\delta
f$ via the freezeout prescription, Eq. (\ref{CF}), we evolve the
system using the second-order viscous hydrodynamic equations
(\ref{BED}) and (\ref{Bshear}) in all the cases.

In Fig. \ref{spectra}, we present the pion transverse-momentum spectra
for the four freezeout conditions discussed above, namely ideal,
first- and second-order Chapman-Enskog, and Grad's 14-moment
approximation. We observe that nonideal freezeout conditions tend to
increase the high-$p_T$ particle production. While the Chapman-Enskog
corrections are small, Grad's 14-moment approximation results in
rather large corrections to the ideal case. This is clearly evident in
the inset where we show the pion yields in the four cases scaled by
the values in the ideal case. These features can be easily understood
from Eqs. (\ref{FCF}) and (\ref{GCF}): The first-order Chapman-Enskog
correction is essentially linear in $p_T$ whereas that due to Grad is
quadratic. The second-order Chapman-Enskog correction is small,
indicating rapid convergence of the expansion up to second order.

\section{HBT radii}

HBT interferometry provides a powerful tool to unravel the space-time
structure of the particle-emitting sources in heavy-ion collisions,
because of its ability to measure source sizes, lifetimes, and
particle emission durations \cite{Lisa:2005dd}. The source function,
$S(x,K)$, for on-shell particle emission is defined such that it
satisfies
\begin{equation}\label{DSF}
\frac{dN}{d^2K_Tdy} \equiv \int d^4x\, S(x,K).
\end{equation}
By comparing the above equation with Eq. (\ref{CF}), we see that the
source function is restricted to the freezeout hypersurface and is
given by
\begin{equation}\label{SF}
S(x,K) = \frac{g}{(2\pi)^3} \int p_\mu d\Sigma^\mu(x') f(x',p) \delta^4(x-x').
\end{equation}
At relatively small momenta, certain space-time variances of the
source function can be obtained, to a good approximation, from the
correlation between particle pairs \cite{Wiedemann:1999qn}.
Space-time averages with respect to the source function are defined as
\begin{equation}\label{ASF}
\mean{\alpha}_K \equiv \frac{\int d^4x\, S(x,K)\alpha}{\int d^4x\, S(x,K)}
=\frac{\int K_\mu d\Sigma^\mu f(x,K)\alpha}{\int K_\mu d\Sigma^\mu f(x,K)},
\end{equation}
where $K_\mu$ is the pair four-momentum.

The longitudinal HBT radius, $R_L$, is calculated in terms of the
transverse momentum, $K_T$, of the identical-particle pair
\cite{Wiedemann:1999qn}:
\begin{equation}\label{HBT}
R_L^2(K_T) = \frac{\int K_\mu d\Sigma^\mu f(x,K)z^2}{\int K_\mu d\Sigma^\mu f(x,K)}.
\end{equation}
In the central-rapidity region, the pair four-momentum is given by
$K^\mu=(K^\tau, K^r, K^\varphi, K^{\eta_s})=(m_T,K_T,0,0)$. The
integration measure is given by $K_\mu d\Sigma^\mu = m_T \cosh(\eta_s)
\tau d\eta_s r dr d\varphi$ with $m_T=\sqrt{K_T^2+m_p^2}$, $m_p$
being the particle mass. Using the relation $z=\tau\,\sinh(\eta_s)$, we
get
\begin{align}\label{HBTB}
 R_L^2(K_T) &= \tau^2\left[\frac{\int K_\mu d\Sigma^\mu f(x,K){\cosh^2(\eta_s)}}
{\int K_\mu d\Sigma^\mu f(x,K)}-1\right], \nonumber \\
&\equiv \tau^2\left[\frac{N[f]}
{D[f]}-1\right].
\end{align}
Note that the integral, $D[f]$, in the denominator in the above
equation is the same as that occurring in the Cooper-Frye prescription
for particle production, Eq. (\ref{CF}), and was already calculated in
the previous section. We next calculate the integral, $N[f]$, in the
numerator.

In the ideal case, $f=f_0$, we have
\begin{equation}\label{Nf0}
N[f_0] = \frac{2 A_\perp \tau z_m}{4\beta}\left(K_3+3K_1\right).
\end{equation}
This leads to the well-known result of Hermann and Bertsch 
\cite{Herrmann:1994rr}
\begin{equation}\label{IHBT}
(R_L^2)^{(0)} = \frac{\tau^2}{z_m }\, \frac{K_2}{K_1},
\end{equation}
which for large values of $z_m$ results in the Makhlin-Sinyukov
formula $(R_L^2)^{(0)} = \tau^2 T / m_T$
\cite{Makhlin:1987gm,Csorgo:1995bi}. Thus in the ideal case,
$(R_L)^{(0)}$ exhibits the so-called $1/\sqrt{m_T}$ scaling.

The first-order calculation requires $N[\delta f_1]$, which is given by
\begin{align}\label{Ndf1}
N[\delta f_1] = \frac{2 A_\perp \tau \Phi}{16\beta\beta_\pi}\Big[\left(2z_p^2+z_m^2\right)K_0 + 2z_p^2K_2 
- z_m^2K_4\Big].
\end{align}
The second-order calculation requires $N[\delta f_2]$, which is given
in the Appendix B. For comparison we also calculate $R_L$ in Grad's
14-moment approximation. This requires $N[\delta f_G]$, which we obtain
as
\begin{align}\label{NdfG}
N[\delta f_G] =&~ \frac{2 A_\perp \tau\Phi z_m}{160\beta\beta_\pi}\Big[\left(2z_p^2-6z_m^2\right)K_1 \nonumber\\
&+ \left(2z_p^2-z_m^2\right)K_3 - z_m^2K_5\Big].
\end{align}
 
In the following, we show that the viscous correction to $R_L$ due to
Grad's 14-moment approximation violates the experimentally observed
$1/\sqrt{m_T}$ scaling 
\cite{Beker:1994qv,Bearden:1998aq,Bearden:2001sy,Adcox:2002uc,LopezNoriega:2002tn}, 
whereas it
is preserved in the Chapman-Enskog case. To this end, we calculate the
first-order viscous correction to $R_L$ in both the cases. Expanding the
$R_L$ in Eq. (\ref{HBT}) to first order in $\delta f$ and using the
relation $z=\tau \sinh(\eta_s)$ we obtain the ideal contribution
\begin{align}\label{RL0}
(R_L^2)^{(0)} =&~  \frac{\int K^\mu d\Sigma_\mu\, f_0 \,\tau^2 \sinh^2(\eta_s)}
{\int K^\mu d\Sigma_{\mu}\, f_0}, 
\end{align}
and the first viscous correction in the two cases
\begin{align}\label{DRL1G}
\left(\delta R_L^2\right)^{(1,G)} =& -(R_L^2)^{(0)} \left(  
\frac{dN^{(1,G)}}{d^2K_T}-\frac{dN^{(0)}}{d^2K_T} \right)\!\!\Big/
\frac{dN^{(0)}}{d^2K_T} \nonumber\\ 
&+ \frac{ \int K^\mu d\Sigma_\mu\,\tau^2\sinh^2(\eta_s)\,\delta f_{1,G} } 
{ \int K^\mu d\Sigma_\mu \, f_0 }.
\end{align}
The ideal radius $(R_L^2)^{(0)}$ was obtained in Eq. (\ref{IHBT}).
Viscous corrections due to the Chapman-Enskog method and Grad's
14-moment approximation can be obtained similarly. By substituting the
viscous correction, $\delta f_1$, from Eq. (\ref {SOVC}) into
Eq. (\ref{DRL1G}), using the results for the particle spectra,
Eqs. (\ref{ICF}) and (\ref{FCF}), and the ideal radius, Eq. (\ref{IHBT}), and
performing the $\eta_s$ integrals, we obtain
\begin{equation}\label{DRL1}
\frac{\left(\delta R_L^2\right)^{(1)}}{\left(R_L^2\right)^{(0)}} = -\frac{\Phi}{16\beta_\pi}\left[16 
+ \frac{4z_p^2}{z_m}\left(\frac{K_0}{K_1}-\frac{K_1}{K_2}\right)\right]. 
\end{equation}
Similarly, for Grad's approximation, Eq. (\ref{CG}), we obtain
\begin{equation}\label{DRLG}
\frac{\left(\delta R_L^2\right)^{(G)}}{\left(R_L^2\right)^{(0)}} = -\frac{\Phi}{20\beta_\pi}
\left[ 20 - 2z_m\left(\frac{K_0}{K_1}-\frac{K_1}{K_2}\right) + 4z_m\frac{K_1}{K_2} \right]. 
\end{equation}

Using the asymptotic expansion of modified Bessel functions of the
second kind \cite{abramowitz},
\begin{equation}\label{ASYMP}
K_n(z_m) = \left(\frac{\pi}{2z_m}\right)^{\frac{1}{2}}e^{-z_m}\left[1+\frac{4n^2-1}{8z_m}+\cdots\right],
\end{equation}
for large $z_m$, we have
\begin{equation}\label{IDENTY}
\frac{K_0}{K_1}-\frac{K_1}{K_2} = \frac{1}{z_m} + {\mathcal O}\left(\frac{1}{z_m^2}\right).
\end{equation}
Hence, for large values of $z_m$, we find
\begin{align}
\left(\delta R_L^2\right)^{(1)} &= -\frac{5\tau^2T\Phi}{4\beta_\pi m_T},
\label{ASYMPRL1}\\
\left(\delta R_L^2\right)^{(G)} &= -\frac{\tau^2T\Phi}{5\beta_\pi m_T}\left( 3+\frac{m_T}{T} \right).
\label{ASYMPRLG}
\end{align}
It is clear from the above two equations that the viscous correction
to $R_L$ in the Chapman-Enskog case preserves the $1/\sqrt{m_T}$
scaling, whereas in Grad's 14-moment approximation it grows as
$m_T/T$ relative to the ideal result, and thus violates the scaling \cite{Teaney:2003kp}.

 \begin{figure}[t]
 \begin{center}
 \includegraphics[scale=0.41]{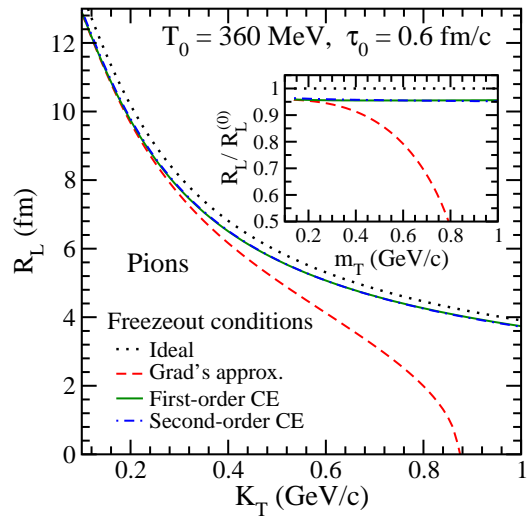}
 \end{center}
 \vspace{-0.4cm}
 \caption{(Color online) Longitudinal HBT radius as a function of the
   transverse momentum $K_T$ of the pion pair, obtained with the
   second-order hydrodynamic evolution, followed by freezeout in
   various scenarios: ideal, Grad's 14-moment approximation, and
   first- and second-order Chapman-Enskog. Inset: HBT radius in
   the above cases scaled by the corresponding values in the ideal case.}
 \label{fhbt}
 \end{figure}

Results for the longitudinal HBT radius, $R_L$, for identical-pion
pairs in central Au-Au collisions, for the four cases discussed above,
are displayed in Fig. \ref{fhbt}. We note that while there is no
noticeable difference between first- and second-order Chapman-Enskog
results compared to the ideal case, they predict a slightly smaller
value for $R_L$. On the other hand, $R_L$ corresponding to Grad's
approximation exhibits a qualitatively different behavior and even
becomes imaginary for $K_T \gtrsim 0.9$ GeV/$c$, which is clearly
unphysical. More importantly, the ratio $R_L/R_L^{(0)}$ shown in the
inset of Fig. \ref{fhbt} illustrates that the $1/\sqrt{m_T}$ scaling,
which is violated in Grad's approximation, survives in the
Chapman-Enskog case.

\section{Summary and Conclusions}

We derived the form of the viscous correction to the equilibrium
distribution function, up to second order in gradients, by employing a
Chapman-Enskog-like iterative solution of the Boltzmann equation in
the relaxation-time approximation. This approach is in accordance with
the formulation of hydrodynamics, which is also a gradient expansion.
We used this form of the viscous correction to calculate the hadronic
transverse-momentum spectra and longitudinal Hanbury-Brown-Twiss
radii and compared them with those obtained in Grad's 14-moment
approximation within the one-dimensional scaling expansion. These
results demonstrate the rapid convergence of the Chapman-Enskog
expansion up to second order, and thus it is sufficient to retain only
the first-order correction in the freezeout prescription. We found
that the Chapman-Enskog method results in softer hadron spectra
compared with Grad's approximation. We further showed that the
experimentally observed $1/\sqrt{m_T}$ scaling of HBT radii, which is
also seen in the ideal freezeout calculation, is maintained in the
Chapman-Enskog method. In contrast, the Grad's 14-moment approximation
leads to the violation of this scaling as well as an imaginary value
for $R_L$ at large momenta. For initial conditions typical of
heavy-ion collisions at the Large Hadron Collider ($T_0 = 500$ MeV and
$\tau_0=0.4$ fm/$c$), we have found that the above conclusions remain
unchanged.

We conclude by recalling the well-known form of the viscous correction
due to Grad's 14-moment approximation,
\begin{equation}
 \delta f_G = \frac{f_0 \tilde f_0}{2(\epsilon+P)T^2}\, p^\alpha p^\beta \pi_{\alpha\beta},
\end{equation}
and the alternate form due to Chapman-Enskog method proposed here,
\begin{equation}
 \delta f_{CE}= \frac{5 f_0 \tilde f_0}{8 P T(u \!\cdot\! p)}\, p^\alpha p^\beta \pi_{\alpha\beta},
\end{equation}
where $\tilde f_0 \equiv 1-r f_0$, with $r=1,-1,0$ for Fermi, Bose, and
Boltzmann gases, respectively. In view of the arguments presented in
this paper, we advocate that the form of $\delta f_{CE}$ proposed here
should be a better alternative for hydrodynamic modeling of
relativistic heavy-ion collisions.

\appendix

\section{CONSTRAINTS ON THE VISCOUS CORRECTION TO THE DISTRIBUTION FUNCTION}

\label{Appendix A}

In this appendix, we show that the form of the viscous correction to
the distribution function, $\delta f$, given in Eq. (\ref {SOVC})
satisfies the matching condition $\epsilon =\epsilon_0$ and the Landau
frame definition $u_\nu T^{\mu \nu} = \epsilon u^\mu$, at each order
in gradients \cite{deGroot}. We also show that $\delta f$ is
consistent with the definition of the shear stress tensor,
Eq. (\ref{FSE}).

The first- and second-order viscous corrections to the distribution
function can be written separately using Eq. (\ref {SOVC}). The
first-order correction is given by
\begin{equation}\label{deltaf1}
\delta f_1 =  \frac{f_0\beta}{2\beta_\pi(u\!\cdot\!p)}\, p^\alpha p^\beta \pi_{\alpha\beta},
\end{equation}
whereas the second-order correction is
\begin{align}\label{deltaf2}
\delta f_2 \!= & -\frac{f_0\beta}{\beta_\pi} \bigg[\frac{\tau_\pi}{u\!\cdot\!p}\, p^\alpha p^\beta \pi^\gamma_\alpha\, \omega_{\beta\gamma} 
-\frac{5}{14\beta_\pi (u\!\cdot\!p)}\, p^\alpha p^\beta \pi^\gamma_\alpha\, \pi_{\beta\gamma} \nonumber\\
&+\!\frac{\tau_\pi}{3(u\!\cdot\!p)}p^\alpha p^\beta \pi_{\alpha\beta}\theta  
-\frac{6\tau_\pi}{5} p^\alpha\dot u^\beta\pi_{\alpha\beta}
+\!\frac{(u\!\cdot\!p)}{70\beta_\pi}\pi^{\alpha\beta}\pi_{\alpha\beta}\nonumber\\
&+\frac{\tau_\pi}{5} p^\alpha\! \left(\!\nabla^\beta\pi_{\alpha\beta}\!\right)\!
-\frac{3\tau_\pi}{(u\!\cdot\!p)^2}\, p^\alpha p^\beta p^\gamma \pi_{\alpha\beta}\dot u_\gamma \!+ \frac{\tau_\pi}{2(u\!\cdot\!p)^2}\nonumber\\
&\times\! p^\alpha p^\beta p^\gamma\! \left(\nabla_\gamma\pi_{\alpha\beta}\!\right) 
-\frac{\beta\!+\!(u\!\cdot\!p)^{-1}}{4(u\!\cdot\!p)^2\beta_\pi} \!\left(p^\alpha p^\beta \pi_{\alpha\beta}\!\right)^{\!2}\!\bigg].
\end{align}

In the following, we show that the $\delta f_i$ given in Eqs. (\ref
{deltaf1}) and (\ref{deltaf2}) satisfies the conditions
\begin{equation}\label{Landau1}
L_1[\delta f_i] \equiv \int dp\, (u\cdot p)^2\, \delta f_i = 0,
\end{equation}
corresponding to $\epsilon =\epsilon_0$, and
\begin{equation}\label{Landau2}
L_2[\delta f_i] \equiv \int dp\, \Delta_{\mu\alpha}u_\beta\, p^\alpha p^\beta\, \delta f_i = 0,
\end{equation}
corresponding to $u_\nu T^{\mu \nu} = \epsilon u^\mu$.

At first order, we obtain
\begin{equation}\label{Landaudf1}
L_1[\delta f_1] = \frac{\beta}{2\beta_\pi}\pi_{\alpha\beta}u_\gamma I_{(0)}^{\alpha\beta\gamma},\quad\!
L_2[\delta f_1] = \frac{\beta}{2\beta_\pi}\pi_{\alpha\beta}\Delta_{\mu\gamma} I_{(0)}^{\alpha\beta\gamma},
\end{equation}
where we define the integral
\begin{equation}\label{mom_int}
I^{\mu_1\mu_2\cdots\mu_n}_{(r)} \equiv \int \frac{dp}{(u\!\cdot\! p)^r} p^{\mu_1}p^{\mu_2} \cdots p^{\mu_n} f_0. 
\end{equation}
The above momentum integral can be decomposed into hydrodynamic tensor
degrees of freedom as
\begin{align}\label{tens_decm}
I^{\mu_1\mu_2\cdots\mu_n}_{(r)} =\ & I_{n0}^{(r)} u^{\mu_1}u^{\mu_2}\cdots u^{\mu_n} 
+ I_{n1}^{(r)} \big(\Delta^{\mu_1\mu_2} u^{\mu_3} \cdots u^{\mu_n} \nonumber\\
&+ \mathrm{perms}\big) + \cdots, 
\end{align}
where we readily identify $I_{20}^{(0)}=\epsilon$ and
$I_{21}^{(0)}=-P$. Using the above tensor decomposition for
$I_{(0)}^{\alpha\beta\gamma}$ in Eq. (\ref{Landaudf1}), we obtain
\begin{equation}\label{Landau1df1F}
L_1[\delta f_1] = 0,\quad L_2[\delta f_1] = 0.
\end{equation}

Similarly, for second-order corrections given in Eq. (\ref{deltaf2}),
we obtain
\begin{align}\label{Landau1df2}
L_1[\delta f_2] =&\, 0 + \!\frac{5\beta}{14\beta_\pi^2}\pi_{\alpha\beta}\pi^{\alpha\beta}\!I_{31}^{(0)}\! 
+ 0 + 0 - \!\frac{\beta}{70\beta_\pi^2}\pi_{\alpha\beta}\pi^{\alpha\beta}\!I_{30}^{(0)} \nonumber\\
& - \frac{\beta\tau_\pi}{5\beta_\pi}(\nabla^\alpha\pi_{\alpha\beta})I_{30}^{(0)} u^\beta  + 0
- \frac{\beta\tau_\pi}{\beta_\pi}(\nabla_\gamma\pi_{\alpha\beta})I_{31}^{(0)} \nonumber\\
& \times\! u^{(\alpha}\Delta^{\beta)\gamma} + \frac{\beta}{2\beta_\pi^2}\pi_{\alpha\beta}\pi^{\alpha\beta}
\!\left(\beta I_{42}^{(0)} + I_{42}^{(1)}\right).
\end{align}
Using the identities 
\begin{align}
I_{nq}^{(r)}&=-\frac{1}{2q+1}I_{n-1,q-1}^{(r-1)}, 
\label{prop1} \\
I_{nq}^{(0)}&= \frac{1}{\beta}\left[-I_{n-1,q-1}^{(0)} + (n-2q)I_{n-1,q}^{(0)} \right],
\label{prop2}
\end{align}
and Eq. (\ref{FOE}), we obtain
\begin{align}\label{Landau1df2f}
L_1[\delta f_2] =&\, -\frac{25}{14\beta_\pi}\pi_{\alpha\beta}\pi^{\alpha\beta} \!
- \frac{3}{14\beta_\pi}\pi_{\alpha\beta}\pi^{\alpha\beta}\!
+ \frac{12}{8\beta_\pi}\pi_{\alpha\beta}\pi^{\alpha\beta} \nonumber\\
&- \frac{5}{2\beta_\pi}\pi_{\alpha\beta}\pi^{\alpha\beta} 
+ \frac{3}{\beta_\pi}\pi_{\alpha\beta}\pi^{\alpha\beta} \nonumber\\
=&\, 0.
\end{align}

A similar calculation leads to
\begin{align}\label{Landau2df2}
L_2[\delta f_2] =&\, 0 + 0 + 0 + \frac{6\beta\tau_\pi}{5\beta_\pi}I_{31}^{(0)}\Delta_\mu^\alpha\dot u^\beta\pi_{\alpha\beta} + 0 \nonumber\\
&- \frac{\beta\tau_\pi}{5\beta_\pi}I_{31}^{(0)}\Delta_\mu^\alpha \left(\!\nabla^\beta\pi_{\alpha\beta}\!\right)
-\frac{6\beta\tau_\pi}{5\beta_\pi}I_{31}^{(0)}\Delta_\mu^\alpha\dot u^\beta\pi_{\alpha\beta} \nonumber\\
& - \frac{\beta\tau_\pi}{\beta_\pi}I_{42}^{(1)}\Delta_\mu^\alpha \left(\!\nabla^\beta\pi_{\alpha\beta}\!\right) + 0 \nonumber\\
=&\, 0.
\end{align}
To obtain the second equality, we have used Eq. (\ref{prop1}) to
replace $I_{42}^{(1)}=-I_{31}^{(0)}/5$.

Next we show that the form of the viscous correction to the
distribution function, $\delta f=\delta f_1+\delta f_2$ given in Eqs. (\ref
{deltaf1}) and (\ref{deltaf2}),
is consistent with the definition of the shear stress tensor given in
Eq. (\ref {FSE}). In other words, we show that $\pi^{\mu\nu}=L_3[\delta
f_1]+L_3[\delta f_2]$, where
\begin{equation}\label{CheckDefn}
L_3[\delta f_i] \equiv \Delta^{\mu\nu}_{\alpha\beta} \int dp \, p^\alpha p^\beta\, \delta f_i.
\end{equation}
 At first order, we get
\begin{equation}\label{CheckDefn1}
L_3[\delta f_1] =\frac{\beta}{2\beta_\pi}\,\Delta^{\mu\nu}_{\alpha\beta}\,\pi_{\gamma\delta}\,I_{(1)}^{\alpha\beta\gamma\delta}.
\end{equation}
Using the tensor decomposition for
$I_{(1)}^{\alpha\beta\gamma\delta}$ in the above equation, we
obtain
\begin{equation}\label{CheckDefn1f}
L_3[\delta f_1] =\frac{\beta}{\beta_\pi}\,I^{(1)}_{42}\,\pi^{\mu\nu} = \pi^{\mu\nu}.
\end{equation}
Here we have used $I^{(1)}_{42}=\beta_\pi/\beta$, obtained 
by employing the recursion relations, Eqs. (\ref{prop1}) and (\ref{prop2}).

Similarly, for the second-order correction $\delta f_2$ given in Eq.
(\ref{deltaf2}), we obtain
\begin{align}\label{CheckDefn2}
L_3[\delta f_2] =&\, -2\tau_\pi \pi_\gamma^{\langle\mu} \omega^{\nu\rangle\gamma}
+ \frac{5}{7\beta_\pi}\pi_\gamma^{\langle\mu} \pi^{\nu\rangle\gamma}
-\frac{2}{3}\tau_\pi \pi^{\mu\nu}\theta + 0 \nonumber\\
&+ 0 + 0 + 0 + \Big( \frac{1}{\beta_\pi}\pi_\gamma^{\langle\mu} \pi^{\nu\rangle\gamma}
+ 2\tau_\pi \pi_\gamma^{\langle\mu} \omega^{\nu\rangle\gamma} \nonumber\\
&+ \frac{2}{3}\tau_\pi \pi^{\mu\nu}\theta \Big)
-\frac{12}{7\beta_\pi}\pi_\gamma^{\langle\mu} \pi^{\nu\rangle\gamma} \nonumber\\
=&\ 0.
\end{align}
Hence $L_3[\delta f]=L_3[\delta f_1]+L_3[\delta f_2]=\pi^{\mu\nu}$.
This result was expected because no second-order term (e.g., $\pi \pi$,
$\pi \omega$, etc.) or their linear combinations, when substituted in
Eq. (\ref{FSE}), can result in a first-order term ($\pi$) which we
have on the left-hand side of Eq. (\ref{FSE}). In fact, each higher-order
correction ($\delta f_n$) when substituted in Eq. (\ref{FSE}) will
vanish. The fact that $\delta f$ given in Eq. (\ref{SOVC}) satisfies the constraints,
as demonstrated in this Appendix, shows that our method of obtaining
the viscous corrections to the distribution function is quite robust.

\section{SECOND-ORDER VISCOUS CORRECTIONS TO HADRON SPECTRA AND HBT RADII}

\label{Appendix B}

Within the one-dimensional scaling expansion, $\dot u = 0 =
\omega^{\mu\nu}$, which reduces the number of terms in
Eq. (\ref{deltaf2}). The nonvanishing terms can be simplified using
Eq. (\ref{visc-fcts}) as
\begin{align}
\delta f_2 \!=&\ \frac{f_0 \beta}{\beta_\pi} \Bigg[-\frac{5\Phi^2 m_T\left\{p_T^2/(4m_T^2) + \sinh^2(y-\eta_s)\right\}}
{14\beta_\pi \cosh(y-\eta_s)} \nonumber \\
&-\frac{\tau_\pi\Phi\, m_T\left\{p_T^2/(2 m_T^2) - \sinh^2(y-\eta_s)\right\}}{3\tau\cosh(y-\eta_s)} \nonumber \\
&- \frac{3\Phi^2 m_T \cosh(y-\eta_s)}{140\beta_\pi} + \frac{\tau_\pi\Phi\, m_T \cosh(y-\eta_s)}{5 \tau}\nonumber \\
&-\frac{\tau_\pi  \Phi\, m_T \sinh^2(y-\eta_s)}{\tau \cosh(y-\eta_s)} 
+ \frac{\Phi^2\beta}{4\beta_\pi \cosh^2(y-\eta_s)}\nonumber \\
&\times\! \left\{\! 1 + \frac{(\beta m_T)^{-1}}{ \cosh(y-\eta_s)} \!\right\}\!\! 
\left\{\!\frac{p_T^2}{2m_T^2} - \sinh^2(y-\eta_s)\!\right\}^{\!2}\Bigg]. 
\label{df2}
\end{align}

The contribution to the hadronic spectra resulting from these
second-order terms is calculated using Eq. (\ref{CF}) as
\begin{align}
\frac{\delta dN^{(2)}}{d^2p_Tdy} \equiv &\ \frac{g}{(2\pi)^3} \int m_T \cosh(y-\eta_s) \tau d\eta_s rdr d\varphi\, \delta f_2\nonumber \\
=&\ \frac{g\,\tau\, A_\perp}{4\pi^3 \beta\beta_\pi} \Bigg[ 
\frac{-5\Phi^2}{56\beta_\pi} \left( z_p^2\,K_0 + 4z_m\,K_1\right) \nonumber \\
& -\frac{\Phi \tau_\pi}{6\tau}\!\left( z_p^2K_0 \!- 2z_mK_1\right)
 \!-\frac{3\Phi^2 z_m^2}{280\beta_\pi}\!\left( K_0 \!+\! K_2\right) \nonumber \\
& +\frac{\Phi \tau_\pi z_m^2}{10\tau}\left( K_0 + K_2 \right)
 -\frac{\Phi\tau_\pi z_m}{\tau} K_1 + \frac{\Phi^2z_m^2}{4\beta_\pi} \nonumber \\
&\times\! \Big\{ z_m X^2 {\mathcal I}_1 - 2 z_m X  K_1  
 + \frac{z_m}{4}\left( K_3 + 3K_1\right) \nonumber \\ 
&  + X^2 {\mathcal I}_2 - 2 X K_0 + \frac{1}{2}\left( K_0 + K_2 \right)\Big\}
 \Bigg]\label{SCF},
\end{align}
where $X\equiv z_p^2/(2z_m^2)+1$, $K_n(z_m)$ are the modified Bessel
functions of the second kind
\begin{equation}
  K_n(z) \equiv \int_0^\infty dt\, e^{-z\cosh (t)} \cosh (nt),
\end{equation}
 and ${\mathcal I}_n$ are the integrals defined as
\begin{equation}
 {\mathcal I}_n(z) \equiv \int_0^\infty dt\, e^{-z\cosh (t)}\, {\rm sech}^n(t),
\end{equation}
with the following properties:
\begin{equation}
 \frac{d^n {\mathcal I}_n(z)}{dz^n} = (-1)^n\,K_0(z), \quad {\mathcal I}_0(z)=K_0(z).
\end{equation}
The expression for hadron spectra up to second order, by setting 
$f=f_0+\delta f_1+\delta f_2$ in the freezeout prescription, Eq. 
(\ref{CF}), becomes
\begin{equation}\label{SCFF}
\frac{d N^{(2)}}{d^2p_Tdy} = \frac{d N^{(1)}}{d^2p_Tdy} + \frac{\delta dN^{(2)}}{d^2p_Tdy}.
\end{equation}

Similarly, within the Bjorken model, one can calculate the
longitudinal HBT radii by including the second-order viscous
corrections in Eq. (\ref{HBTB}) using Eq. (\ref{df2}). To this end, we
calculate $N[\delta f_2]$ by setting $f=f_0+\delta f_1+\delta f_2$ in
Eq. (\ref{HBTB}) and performing the integrations
\begin{align}
 N[\delta f_2] =&\ \int m_T \cosh^3(y-\eta_s) \tau d\eta_s rdr d\varphi\, \delta f_2 \nonumber \\
=&\ \frac{2 A_\perp \tau}{\beta\beta_\pi} \Bigg[ 
\frac{-5\Phi^2}{112\beta_\pi} \Big\{ \left(z_p^2-z_m^2\right)\,K_0 + z_p^2\,K_2 \nonumber \\ 
& +z_m^2\,K_4\Big\} -\frac{\Phi \tau_\pi}{24\tau}\Big\{ \left(2z_p^2+z_m^2\right)\,K_0 + 2z_p^2\,K_2\nonumber \\
&-z_m^2\,K_4\Big\} -\frac{3\Phi^2 z_m^2}{1120\beta_\pi}\left( 3K_0+4 K_2+K_4\right)\nonumber \\
& + \frac{\Phi\tau_\pi z_m^2}{40\tau} \left( 3K_0+4 K_2+K_4\right) 
 -\frac{\Phi\tau_\pi z_m^2}{8\tau} \big( K_4 \nonumber \\
&-K_0\big) + \frac{\Phi^2z_m^2}{4\beta_\pi} \bigg\{\!\!\left(X^2-X+\frac{3}{8}\right)\!K_0 +\bigg(z_m X^2 \nonumber \\
&-\frac{3}{2}z_mX+\frac{5}{8}z_m\bigg)K_1 + \left(\frac{1}{2}-X\!\right)\!K_2 + \bigg(\frac{5}{16}z_m \nonumber \\
&-\frac{1}{2}z_mX \bigg)K_3 + \frac{1}{8}K_4 + \frac{1}{16}z_mK_5 \bigg\} \Bigg].
\end{align}

\end{document}